\documentstyle[prl,aps,preprint,tighten,floats,aps,amssymb]{revtex}

\def\a{\alpha}

\def\c{\chi}

\def\t{\tau}

\def\bo{{\raise.15ex\hbox{\large$\Box$}}}               
\def\pr{\prod}                                          
\def\face{{\raise.2ex\hbox{$\displaystyle \bigodot$}\mskip-2.2mu \llap {$\ddot
        \smile$}}}                                      

\def\leftrightarrowfill{$\mathsurround=0pt \mathord\leftarrow \mkern-6mu
        \cleaders\hbox{$\mkern-2mu \mathord- \mkern-2mu$}\hfill
        \mkern-6mu \mathord\rightarrow$}       
\def\dvec#1{\vbox{\ialign{##\crcr
        \leftrightarrowfill\crcr\noalign{\kern-1pt\nointerlineskip}
        $\hfil\displaystyle{#1}\hfil$\crcr}}}           

\def\beq{\begin{equation}}
\def\eeq{\end{equation}}

\def\beqx{\begin{displaymath}}
\def\eeqx{\end{displaymath}}

\def\beqa{\begin{eqnarray}}
\def\eeqa{\end{eqnarray}}

\def\pl#1#2#3{Phys.~Lett.~{\bf B {#1}}, #3 (19{#2})}
\def\np#1#2#3{Nucl.~Phys.~{\bf B {#1}}, #3 (19{#2})}

\def\pr#1#2#3{Phys.~Rev.~{\bf D {#1}}, #3 (19{#2})}

\def\ijmp#1#2#3{Int.~J.~Mod.~Phys.~{\bf A {#1}}, #3 (19{#2})}

\catcode`@=11
\def\@citex[#1]#2{\if@filesw\immediate\write\@auxout{\string\citation{#2}}\fi
  \def\@citea{}\@cite{\@for\@citeb:=#2\do
    {\@citea\def\@citea{,\penalty\@m}\@ifundefined
      {b@\@citeb}{{\bf ?}\@warning
       {Citation `\@citeb' on page \thepage \space undefined}}%
\hbox{\csname b@\@citeb\endcsname}}}{#1}}

\def\citer{\@ifnextchar [{\@tempswatrue\@citexr}{\@tempswafalse\@citexr[]}}
\def\@citexr[#1]#2{
  \if@filesw\immediate\write\@auxout{\string\citation{#2}}\fi
  \def\@citea{}\@cite{\@for\@citeb:=#2\do
    {\@citea\def\@citea{-\penalty\@m}\@ifundefined
       {b@\@citeb}{{\bf ?}\@warning
       {Citation `\@citeb' on page \thepage \space undefined}}%
\hbox{\csname b@\@citeb\endcsname}}}{#1}\normalsize}
\catcode`@=12

\begin{document}
\draft
\date{\today}
\preprint{\vbox{\baselineskip=12pt
\rightline{UPR-0867-T}
\vskip0.2truecm
\rightline{FERMILAB-Pub-99/369-T}
\vskip0.2truecm
\rightline{\tt hep-ph/0001073}}}
\title{Alternative Supersymmetric Spectra} 
\author{Lisa Everett$^*$, Paul Langacker$^\dagger$,
Michael Pl\"umacher$^\dagger$, and Jing Wang$^{\dagger,**}$}
\address{
$^*$Department of Physics\\
University of Michigan, Ann Arbor, MI 48109  , USA\\
$^\dagger$
Department of Physics and Astronomy \\ 
University of Pennsylvania, Philadelphia PA 19104-6396, USA\\
${}^{**}$Theoretical Physics Department\\
Fermi National Accelerator Laboratory\\   
Batavia, Illinois, 60510, USA
}
\maketitle

\begin{abstract}
We describe the features of supersymmetric spectra,
alternative to and qualitatively different from that of most
versions of the MSSM.  The spectra are motivated by extensions
of the MSSM with an extra $U(1)'$ gauge symmetry, expected in
many grand unified and superstring models, which provide a
plausible solution to the $\mu$ problem, both for models with
supergravity and for gauge-mediated supersymmetry breaking.
Typically, many or all of the squarks are rather heavy
(larger than one TeV), especially for the first two families,
as are the sleptons in the supergravity models.
However, there is a richer spectrum of
Higgs particles, neutralinos, and (possibly) charginos.  Concrete
examples of such  spectra are presented, and the phenomenological 
implications are briefly discussed.
\end{abstract}
\newpage

\noindent {\it Introduction}. 

The minimal supersymmetric standard
model (MSSM) and its simple
extensions contain many free parameters associated with
supersymmetry breaking.  Most analyses have been based on two
generic classes of models of soft supersymmetry breaking: (1)
supergravity, in which supersymmetry breaking in a hidden sector is
transmitted to the observable sector via supergravity.  One usually
assumes universality or at least a comparable scale for the soft
parameters at the Planck scale.  (2) Gauge-mediated models, with the
breaking transmitted via messenger fields at relatively low energy,
such as $10^5$ GeV.  In both cases the scale of the soft
supersymmetry breaking parameters ultimately sets the electroweak
scale via radiative electroweak breaking, provided that the
supersymmetric $\mu$ parameter is of a comparable magnitude.  
Once universality is relaxed there are many free parameters in
supergravity, and there are many versions of gauge-mediation.  However,
in both cases a typical spectrum involves sparticle masses in the
several hundred GeV range due to naturalness arguments; {\it i.e.}, 
the mass scale of the superpartners should be in this range 
(and at most $\sim {\cal O}(1 \, \rm TeV)$) for SUSY to explain the
origin of the electroweak scale without excessive fine-tuning.  Most
studies of the implications for current and future colliders and
precision measurements have been based on such a spectrum.  

However, it is well known that naturalness does not
necessarily require that all sparticles have masses below the TeV
scale \cite{oldheavy}.  In the scalar sector, naturalness only
constrains the masses of the third generation sfermions and the
electroweak Higgs doublets, as these are the fields which have large
Yukawa couplings and thus play dominant roles in radiative
electroweak symmetry breaking.  Therefore, the sparticle masses of
the first and second generations can be significantly larger
than the other sparticle (and particle) masses without violating
naturalness criteria.  
Recent work has demonstrated that this hierarchy can be generated
dynamically via renormalization group evolution (first pointed out in
\cite{kolda} and investigated in the context of grand
unified models in \cite{bagger}). In this scenario, the soft
supersymmetry breaking scalar mass-squared parameters can be
multi-TeV ($\sim 4$ TeV) at the high
scale (while the gaugino masses and scalar trilinear couplings are
$\sim M_W$; the Higgs and third generation masses are driven to
smaller values due to their large Yukawa couplings, while 
the first and second generations remain heavy. The results of a 
recent extension of this framework including the possibility of
multi-TeV $A$ parameters \cite{bagger2} indicate that such inverted 
hierarchies can be generated with the first two generations up 
to $\sim 20$ TeV. This scenario has distinctive implications (such as in
collider searches for superpartners; see, e.g., \cite{kolda,barger}), and
can be advantageous  phenomenologically, as stringent laboratory
constraints on the SUSY  parameter space from flavor-changing neutral
currents (FCNC) and CP-violation can be considerably weakened
\cite{oldheavy}.   Another possibility in supergravity models pointed
out in \cite{moroi} is that since the Higgs soft mass-squared parameter
at the electroweak scale can be quite insensitive to the initial values of
the scalar masses due to ``focus-point''behavior of the RGE's, 
scalar masses for all three generations of squarks and
sleptons of order $2-3$ TeV can be consistent with naturalness (see 
\cite{agashe} for a discussion in the context of gauge-mediated models).

The purpose of this paper is to point out that there is another class
of (string-motivated) models based on gauge extensions of the MSSM
with an additional $U(1)'$ gauge group  in which this type of
spectrum is naturally achieved.   In these models, it has been shown
that the $U(1)'$ may be broken at the TeV scale by a radiative
mechanism analogous to that for electroweak breaking provided there
are sufficiently large Yukawa couplings of a standard model singlet $S$
which carries $U(1)'$ charge \cite{cl}.  Such extended gauge groups,
exotic particle content, and large Yukawa couplings are generically
present in classes of quasi-realistic perturbative superstring
constructions. These models also provide
an alternative resolution to the $\mu$ problem 
of the MSSM, 
since gauge
invariance can forbid the elementary $\mu$ term while an effective
$\mu$ term can be generated via a trilinear coupling of the SM
singlet to the two electroweak Higgs doublets \cite{cl,suematsu}.  
Furthermore,
the enhanced symmetry avoids the problems of
domain walls, which are common to models involving an effective
$\mu$ generated by the VEV of a scalar but not associated with the
breaking of an extra gauge symmetry \cite{king}.  

In this framework the VEV of the singlet field sets the scale of the
$Z'$ mass. This VEV is generally of order several TeV since the
nonobservation of an additional $Z'$ boson and the stringent
constraints on  the $Z-Z'$ mixing angle $\alpha_{Z-Z'}$ typically
require that the mass of the $Z'$ is significantly heavier than the
$Z$ mass (the lower bounds on $M_{Z'}$ are model dependent, but
are in the range of 500 GeV to 1 TeV or so).
Since the singlet VEV is achieved radiatively, its value generally
sets the scale of the required initial values of the soft breaking
parameters.  Typical supersymmetry breaking parameters are at the TeV
scale,  at least for the first two generations.
However, there is typically a much richer spectrum of Higgs particles
and neutralinos, as well as the $Z'$ and (usually) exotic
fermions and their partners.  Specific models based on perturbative 
heterotic string constructions also involve extended chargino 
sectors \cite{chl5}.  

Since the electroweak and $U(1)'$ symmetry breaking are 
coupled in these models, the large ratio of the $Z'$ and $Z$ masses
requires a certain amount of tuning of the parameters (cancellations
are needed for the expectation values of the Higgs doublets to be
sufficiently small). Nevertheless, such models are worth exploring as 
viable alternatives to the MSSM which are well motivated
theoretically both within quasi-realistic string constructions
and GUT models.  An additional motivation to consider such models
seriously arises from recent precision
electroweak data. The $Z$ lineshape and atomic parity data hint at  the
existence of an extra
$Z'$ at a scale around 1 TeV \cite{erler} \footnote{The implications of the
atomic parity data alone have been considered recently in \cite{recentapv}.
For earlier references, see \cite{erler}.}. 
In this paper, we illustrate typical spectra from several concrete models, some
with supergravity-mediated SUSY soft breaking parameters and another with
gauge-mediated supersymmetry breaking, and comment briefly on phenomenological
implications.  \\

\noindent {\it Results: Alternative Supersymmetric Spectra}.

The models we consider are extensions of the MSSM 
with an additional nonanomalous $U(1)'$ gauge symmetry and additional
matter fields, typically including
both SM singlets (with $U(1)'$  charges) and SM exotics. 
Such models are motivated from a class of quasi-realistic
(perturbative heterotic)   
superstring models \cite{chl5}. It was shown in \cite{chl5} that after
vacuum restabilization this class of string models generically contain
extended Abelian gauge structures and additional matter content 
at the string scale. The trilinear couplings, which can be calculated
exactly in string perturbation theory, usually include the top
quark Yukawa coupling and the coupling between Higgs doublets and a
singlet field, an effective $\mu$-term.
The coefficients of these couplings are calculable in string theory,
and are of ${\cal O}(1)$. A phenomenological analysis of an explicit
four-dimensional string
model with these features \cite{chl5} demonstrated that the large
Yukawa couplings trigger the radiative breaking of the $U(1)'$ by
driving the soft supersymmetry breaking mass-squared parameters of
the SM singlet fields $S_i$ negative at low energies via
renormalization group evolution, as argued on general grounds 
in \cite{cl,ddemir,E6}.
For supergravity models with such large Yukawa couplings the $U(1)'$
breaking will either be at the electroweak scale (i.e., up to a TeV or so) 
\cite{ddemir} or at a large intermediate scale \cite{intermediate}
if the breaking occurs along a $D$-flat direction (we do not consider
the intermediate scale $Z'$ case further in this paper).
Electroweak scale breaking can also be implemented in models with
gauge-mediated supersymmetry breaking \cite{gmsb,fnal}.

A general analysis of these scenarios in the supergravity-mediated
supersymmetry breaking framework was analyzed in a minimal
model with no additional exotics \cite{ddemir}. However, this model
is not $U(1)'$ anomaly free, and thus does not have the necessary
ingredients for a fully realistic theory, and thus the case of a  
string-motivated and anomaly-free $E_6$-type model   was 
proposed in \cite{ddemir} and studied in \cite{E6}. These analyses 
demonstrated that there are corners of parameter space for which a  
phenomenologically acceptable $Z-Z'$ hierarchy at the electroweak
scale can be obtained. In these scenarios, the $U(1)'$ breaking is
radiative and triggered by a large ${\cal O}$(TeV) SM-singlet VEV
\footnote{Another possibility is that the symmetry breaking is
driven by  a large value of the soft
supersymmetry breaking trilinear coupling \cite{ddemir}. However, this solution
yields a light $Z'$ that is phenomenologically excluded except in the 
case of models with certain (leptophobic) couplings, and thus we do
not consider this scenario further in this paper.}. This solution
provides a $Z'$ mass close to the natural upper limit of $1-2$ TeV,
with the electroweak scale achieved via cancellations that require a
certain amount of tuning of the soft mass parameters. This is the
least desirable feature of these models.

In all these models, the low energy spectra displayed features that are
different from the standard MSSM spectrum. In general, the
requirement of a phenomenologically acceptable $Z-Z'$ hierarchy leads
to low energy values of some or all of the scalar masses that are
generically a few TeV. This feature can be understood heuristically
within the supergravity-mediated supersymmetry breaking scenarios,
where the boundary conditions are implemented at the string scale
$\sim M_{G}$~\footnote{The small discrepancy between the observed
unification scale $M_G$ and the perturbative heterotic string  scale
is not significant for the cases considered here.} as follows.  In
the limit of $\langle S \rangle \gg \langle H_1 \rangle, \langle H_2
\rangle $, the  $Z'$ mass
is set by the soft mass-squared parameter $m_S^2$ of the singlet
$S$ at low energies by:  $M_{Z'} \sim \sqrt{-2m_S^2}$ (in
this limit the $U(1)'$ breaking can be considered separately from
the electroweak breaking). To obtain the large and negative $m_S^2$
parameter at low energies and to
avoid large fine-tuning, in  general it is necessary that the singlet
couples with ${\cal O}(1)$ Yukawa couplings to additional SM exotic
quarks, as such couplings drive the singlet mass-squared parameter
strongly to negative values.
In this case, the RG evolution provides the desired value of $m_S^2$
at low energies provided that the scale of the soft mass-squares of the exotic
quarks at $M_{G}$ is about a few TeV. This scale  determines the
typical magnitudes of the soft mass-squared parameters in such
phenomenologically acceptable models (which are not excessively tuned
either at the electroweak scale or at the high scale). 
The first and second generation sparticles (whose soft mass-squared
parameters do not run significantly due to their smaller Yukawa
couplings) are typically a few TeV.
However, a certain amount of fine-tuning is needed to obtain low
energy values of the Higgs soft mass-squared parameters of the order
of the electroweak scale.
Since the $U(1)'$ symmetry breaking is at the TeV scale, 
there are also additional Higgs bosons and neutralinos in 
the low energy spectrum. In the large $\langle S \rangle$ limit,
some of these additional states acquire masses $\sim M_{Z'}$. 
We note that in the explicit string-derived model  analyzed in
\cite{chl5}, the couplings are more complicated, and an extended
Higgs sector must be invoked to achieve a realistic $Z-Z'$ hierarchy.
As a result, the low energy spectrum includes additional charginos as
well as Higgs bosons and neutralinos.

To illustrate these features, we now
turn to several supergravity models \cite{ddemir,E6} and demonstrate
the symmetry breaking pattern and the low energy spectrum explicitly.
For the sake of simplicity, the models discussed are those with a
minimal Higgs sector of two electroweak doublets and one SM
singlet.
For each model, we display the relevant
mass parameters at both the electroweak scale and the string (or GUT)
scale in Table I and II, and present the detailed low energy spectra
explicitly in Table III.

The first example we consider of an anomaly-free model with
$U(1)'$ charges that allow an induced $\mu$ term $(Q_1+Q_2+Q_S=0)$,
where $Q_1$, $Q_2$, and $Q_S$ are respectively the $U(1)'$ charges of
$H_1$, $H_2$, and $S$,   and
also does not include additional SM exotics was first presented in
the Appendix of \cite{ddemir}. The charge assignments in this model
(in self-evident notation) are given by 
\begin{equation}
\begin{array}{ll}
{Q_E}_3=Q_2-Q_1, & {Q_L}_3=-Q_2, \vspace{0.1cm}\\
{Q_Q}_3=-\frac{1}{3}Q_1, & Q_S=-(Q_1+Q_2), \vspace{0.1cm}\\
{Q_D}_3=\frac{1}{3}(Q_1+3Q_2), & {Q_U}_3=\frac{1}{3}(Q_1-3Q_2),
\label{ddemircharges}
\end{array}
\end{equation}
for arbitrary $Q_1$ and $Q_2$, and the first and second families have zero
$U(1)'$ charges. We stress that this model is only semi-realistic; while
these charge assignments are consistent with gauge invariance conditions
for the top quark ${Q_U}_3+{Q_Q}_3+Q_2=0$ and the tau lepton 
${Q_E}_3+{Q_L}_3+Q_1=0$ Yukawa interactions, the bottom quark Yukawa
interaction (and those of the first two generations) is forbidden by the
symmetry. The bottom quark mass can be generated from a higher-dimensional
operator, but its value is suppressed by the $U(1)'$ breaking scale and
thus is too small. However, we present this model as a minimal example in
which to display the patterns of the $U(1)'$ symmetry breaking and
resulting spectra. Nonuniversal boundary conditions (or the addition of
exotics) at the string scale are required to drive the singlet
mass-squared parameter negative at the electroweak scale in this model.
The boundary conditions at the string scale are presented in Table I,
for an example in which $M_{Z'}=1$ TeV, the $Z-Z'$ mixing is
$\alpha_{Z-Z'}=2 \times 10^{-3}$, and $\tan \beta=2$.

Another example, which provides acceptable anomaly free
$U(1)'$ quantum numbers and is approximately consistent with gauge
unification, is a string-motivated model with $E_6$ particle
content (without $E_6$-type relations among the Yukawa
couplings) \cite{ddemir,E6}.  The particle content of the model under
consideration includes three $E_{6}$ 27-plets, each of which includes an
ordinary family, two Higgs-type doublets,
two standard model singlets, and two exotic $SU(2)$-singlet quarks with
charge $\pm 1/3$. We also assume a single vector-like  pair of
Higgs-type doublets from a $27+27^{*}$, which does not introduce any
anomalies. The particle content is consistent with gauge unification.
It is
further assumed that  only a subset of these fields (the SM Higgs doublets, SM
singlet
$S$, and an exotic quark pair
$D$ and
$\bar{D}$) play significant roles in the radiative breaking due to the
presence of trilinear superpotential couplings (with ${\cal O}(1)$
coefficients) of the form $S H_{1} \cdot H_2$ and $S D \bar{D}$.  

The $U(1)'$ symmetry breaking patterns of this model were analyzed in
detail in \cite{E6} assuming a general set of supergravity-mediated 
soft supersymmetry breaking mass parameters; we refer the reader to
this work for further details. In general, non-universal boundary
conditions are required to achieve the desired hierarchy. The
soft parameters and low energy spectrum of the first
numerical example of this model ($E_6$(UA)), which includes only the
dominant effect of the top Yukawa couplings and assumes universal
$A$-parameters at $M_{G}$, are presented for a case in which 
 $M_{Z'}=1700$ GeV, $\alpha_{Z-Z'}=2 \times 10^{-3}$.
The second numerical example of this model ($E_6$(NUA)), which
has non-universal $A$-terms and also includes the bottom, tau and
charm Yukawas, is presented for a case in which $M_{Z'}=1600$ GeV,
$\alpha_{Z-Z'}=1 \times 10^{-3}$. The non-universal $A$ parameters
result in a different pattern in the squark spectrum, as presented
in Table III.

This class of models was also analyzed recently assuming
gauge-mediated supersymmetry breaking \cite{gmsb}. The
particle content of the observable sector in the particular example
considered includes the MSSM fields, as well as vector pair of quark
singlets ($D$ and $\bar{D}$), and an additional singlet field whose  
couplings to the two Higgs doublets and to $D$, $\bar{D}$ are allowed
by gauge invariance.
The supersymmetry breaking scale is set
to the standard value of $10^5$ GeV. In this model it is assumed that
the messenger fields are not charged under the $U(1)'$ symmetry;
therefore, the soft mass-square of the singlet field  is zero at
the messenger scale. As a result, to achieve
a desired value of $m_{S}^2$ at the electroweak scale over the     
short period of RGE running (from $10^5$ GeV to $\sim~ 10^2$ GeV), the
soft mass-squares of the scalar  exotic quarks generated at the
messenger scale are required to be of order TeV, which thus sets the
mass scale for the masses of the other squarks.  A numerical example
of the model is presented in Table I and II, with the $Z'$ mass  of
$1110$ GeV and the mixing angle $\alpha_{Z-Z'} = 0.004$.

An inspection of Table III  indicates that the low energy
spectra of each of these models share several common    
features. The scalar particles generally have masses at the TeV
scale, higher than the standard scenarios in the MSSM. In
particular, the squarks of the first and second generation have
masses about $1-3$ TeV in each of the examples. The first $E_6$
example, with universal $A$-parameters at the GUT scale, predicts a  
hierarchy between the third family and the first two family squarks;
in particular the stops and sbottoms are much lighter. In contrast,
in the second $E_6$ example with  non-universal $A$-parameters, the
squarks of the three families can have  masses
of the same order. In the case in which the third family
sparticles are also heavy, there is an additional tuning issue
because the stops enter at the loop level of the Higgs potential;
however, given that cancellations between terms of order $(\rm 
TeV)^2$ already are present at the tree level, the tuning is not
significantly worsened for stops with TeV masses.  
In the gauge-mediated model, the squarks
(including the exotics) acquire TeV scale masses at the
messenger scale; since the running time is very short (from $10^5$
GeV to $\sim 10^2$ GeV), their masses stay  heavy ${\cal O}$(TeV).
In the slepton sector, both of the $E_6$ models predict heavy
sleptons with masses above around one TeV,  while the gauge mediated model
has sleptons of a few hundred GeV due to the gauge coupling   
hierarchy at the messenger scale (see also \cite{agashe}). 

We now turn to a discussion of the Higgs and neutralino sectors,
as discussed in detail in \cite{ddemir}. With the assumption of a
minimal set of Higgs fields, in addition to the MSSM Higgs bosons
there is one additional neutral CP-even Higgs boson which is
predominantly the real component of the singlet field and has a mass
$\sim M_{Z'}$ in the large $\langle S \rangle$ limit. In the
neutralino sector, there are two additional
neutralinos: one extra gaugino (corresponding to the fermionic
partner of the $Z'$) and an extra Higgsino (corresponding to the
fermionic partner of the $S$ field). In the large $\langle S \rangle$
limit, these neutralinos mix and have masses controlled by $M_{Z'}$.
We also note that the upper bound on the tree level mass of the
lightest Higgs receives a contribution from the $U(1)'$
$D$-term    \cite{ddemir} 
and thus can be heavier than that of the MSSM,
which is a particular feature of this class of
models. \\

\noindent {\it Concluding Remarks}.

The purpose of this paper has been to emphasize two main points: (i)
supersymmetric models with an additional $U(1)'$ gauge
symmetry broken at the TeV scale are well motivated extensions of the
MSSM both theoretically and phenomenologically, and (ii)
the characteristic low energy mass spectra of this class of models
exhibit patterns which have distinctive features compared to that of
the MSSM. In particular, the strong phenomenological constraints on
the $Z'$ mass and mixing with the ordinary $Z$ dictate that the
$U(1)'$ is broken by a large singlet VEV of order several TeV, which
sets the initial scale of the soft scalar mass-squared parameters.
The resulting low energy spectra generically have heavy scalars, as
well as a richer spectrum of Higgs bosons, neutralinos, and 
possibly charginos.  In this scenario, the electroweak scale is
generated by cancellations, which in turn suggests a  natural
upper limit of the mass scale of the heavy scalars (and the $Z'$
mass) to be of order several TeV to avoid excessive fine-tuning.

We conclude with a brief discussion of the phenomenological
implications of the mass spectra in this class of models.  In
general, the heavy squarks (and sleptons in the supergravity models)
can lead to
distinctive phenomenological signatures and can ease the strong
constraints on the SUSY parameter space from FCNC and CP violation
(see e.g. \cite{fcnccp,fcnc,cp} and references therein for the
analysis of these processes within the MSSM) as discussed recently in
\cite{oldheavy,barger}. In the models considered here the heavy scalars in these
models are typically in the range $1-3$ TeV; as such,  it is well
known that flavor changing neutral current (FCNC) operators  due to
box diagrams are typically suppressed compared to the MSSM because of
the larger scale, although one must still rely somewhat on the
assumption that the soft scalar mass-squares that are generated due
to some (unknown) supersymmetry breaking mechanism 
are diagonal in flavor space. 
In our RGE analysis, the
scalar mass-squares and the $A$ terms are assumed to be diagonal (but
not necessarily universal).
For this reason, we will not go
into a detailed analysis of the implications of the spectra presented
in our paper on the FCNC and CP-violating processes. Instead, note
that with squark masses of order a few TeV, the requirement of
universality of the soft scalar masses can be relaxed compared to the
case of the MSSM. Namely, the splitting between the scalar masses of
the three quark families $|m_{q_1} - m_{q_2}|/m_{q_3}$ can be as
large as $\cal O$($1$), while still satisfying present experimental
bounds on FCNC. We also point out that in the models considered there can also
be new flavor changing effects due to family non-universal $U(1)'$
couplings (e.g., the string-derived model in \cite{chl5} has family
non-universal $U(1)'$ charges). While such family non-universal
couplings are subject to severe constraints for the couplings of the
first two generations, the third generation couplings are less
constrained. An complete analysis of these effects is currently
underway \cite{pl}.

In addition to the effects from the heavy scalars, the extended
gauge, Higgs, neutralino, and (possibly) chargino sectors implicit in
these models have a number of phenomenological consequences.  The
implications  include new expectations for precision experiments \cite{erler}
and collider searches, and possibly new patterns for dark matter
\cite{jose}.  In addition, the presence of the $U(1)'$ symmetry and
the exotics can have effects on $R$-parity violation, neutrino
masses, quark and charged lepton masses and mixings,
and scenarios for
baryogenesis.   A comprehensive study of such issues
is beyond the scope of this paper and is
deferred for a future study.

\acknowledgments
We thank M. Cveti\v{c}, J. Erler, and J. R. Espinosa for helpful
discussions and suggestions.  This work is supported in part by the
U.S. Department of Energy Grants No.~EY-76-02-3071 (P.L.,M.P.),
DE-FG02-95ER40899 (L.E.), and DE-AC02-76CH03000 (J.W.), and in part by the
Feodor Lynen Program of the Alexander von Humboldt Foundation (M.P.).

\begin{table}
\begin{center}
\begin{tabular}{|l|r r|r r|r r|r r|}
 & \multicolumn{2}{c}{SF} & \multicolumn{2}{c}{$E_6$ (UA)}
 & \multicolumn{2}{c}{$E_6$ (NUA)} &
   \multicolumn{2}{c}{GMSB} \\[1ex]
\hline
$h_t$    &  0.72    & 1.02&
             1.7 &  1.27 &
           0.85 &  1.26 &
           0.84 & 0.98 \\[1ex]
$h_S$    &      0.72 & 0.57& 
             1.7 &  0.21 &
           0.85 &  0.24 &  
           0.47 & 0.40 \\[1ex]
$h_D$    & --    & -- &    
             1.7 &  1.27 &
           0.85 &  1.23 &
             0.70 & 0.84 \\[1ex]
$h_b$    &  --   &--  & 
            --  & --   &
           0.01  &  0.06 &
           0.30 & 0.42 \\[1ex]
$h_c$    &  --     &  --&
             -- & --   &
           $3\times 10^{-3}$ &  0.01 & 
            --  & --  \\[1ex]
$h_{\t}$ &    --   &-- & 
              -- & --    &
           0.01  & 0.02  &   
           0.17 & 0.18 \\[1ex]
$M_1$    &   100    & 41&
             371 &    54 &  
             650 &    95 &   
             147 & 135 \\[1ex]
$M_2$    &   100    & 82&
             371 &   108 &
             650 &   190 &
             279 & 269 \\[1ex]
$M_3$    &   100    & 290&
             371 &   371 &
             650 &   650 & 
             801 & 955 \\[1ex]
$M_1'$   &   100    & 35&
             371 &    52 & 
             650 &    92 & 
             0 & 0 \\[1ex]
$A_Q$    & 1800      & 440 &   
           $-$2960 &   281 &   
           $-$2180 &   494 &
             0 & 449 \\[1ex]
$A_S$    &  2240     & 600 &   
           $-$2960 &  $-$188 &   
            2180 &    50 &
               0 &   $-$53 \\[1ex]
$A_D$    &   --    &  -- &
           $-$2960 &   284 &
            2180 &   801 &
             0 &     494 \\[1ex]
$A_b$    &  --  &  --&   
           $-$2960 & $-$1550 & 
           $-$2180 &   142 & 
             0 & 574\\[1ex]
$A_c$    &   --    & -- &
           $-$2960 & $-$1520 & 
           $-$2180 &   316 &
             --   & --  \\[1ex]
$A_{\t}$ &   --    & -- &
           $-$2960 & $-$2700 &
            2180 &  2410 & 
             0 & 24 \\[1ex]
\end{tabular}
\end{center}
\caption{The Yukawa couplings $h_i$, gaugino masses $M_a$, and the soft
trilinear couplings $A_i$ for each discussed model. SF refers to the model
in eq.~(\ref{ddemircharges}) with $Q_1=-1$, $Q_2=-1/2$. $E_6$ (UA) and
$E_6$ (NUA) are respectively the models with universal and non-universal
$A$ terms. GMSB is the $E_6$ model with gauge mediated supersymmetry
breaking.  $h_{t,b,c,\t}$ are the 
Yukawa couplings for the top, bottom, charm, and tau, respectively. $h_S$
is the Yukawa coupling of the singlet $S$ to the Higgs doublets, and $h_D$
(for the $E_6$ models) is the Yukawa coupling of $S$ to the exotic quark
pair $D,\bar{D}$.  For each model the first column refers
to the parameters at the string or messenger scale, and the second at
the electroweak scale.  For each table, the blank entries refer to
arbitrary quantities that are essentially irrelevant to the symmetry
breaking pattern, while the dashes refer to
small quantities that are neglected or to states that are absent
in the model.}
\end{table}

\begin{table}
\begin{center}
\begin{tabular}{|l|r r|r r|r r|r r|}
 & \multicolumn{2}{c}{SF} & \multicolumn{2}{c}{$E_6$ (UA)} &
   \multicolumn{2}{c}{$E_6$ (NUA)} & \multicolumn{2}{c}{GMSB} \\[1ex]
\hline
$m_1^2$               & $(1150)^2$ & $  (577)^2$ &
                        $(1920)^2$ & $  (302)^2$ &
                        $(1660)^2$ & $ $-$(721)^2$ &
                        $ (351)^2$ & $(83)^2$ \\[1ex]
$m_2^2$               & $(2660)^2$ & $ $-$(302)^2$ &
                        $(5550)^2$ & $  (681)^2$ &
                        $(5940)^2$ & $  (442)^2$ &   
                        $ (351)^2$ & $ $-$(778)^2$ \\[1ex]
$m_S^2$               & $(1220)^2$ & $ $-$(707)^2$ &
                        $(3600)^2$ & $$-$(1210)^2$ &
                        $(3830)^2$ & $$-$(1150)^2$ &
                        $ 0 $      & $ $-$(821)^2$ \\[1ex]
$m_{Q_3}^2$           & $(2730)^2$ & $ (1000)^2$ &
                        $(2920)^2$ & $  (131)^2$ &
                        $(3170)^2$ & $ (1000)^2$ &
                        $(1350)^2$ & $ (1410)^2$ \\[1ex]
$m_{u_3}^2$           & $(2250)^2$ & $ (1000)^2$ &
                        $(4180)^2$ & $  (181)^2$ &
                        $(4550)^2$ & $ (1010)^2$ &
                        $(1310)^2$ & $ (1290)^2$ \\[1ex]
$m_D^2$               & $-$     & $-$    &
                        $(1630)^2$ & $  (993)^2$ &
                        $(1630)^2$ & $ (1000)^2$ &   
                        $(1310)^2$     & $(1440)^2$ \\[1ex]
$m_{\bar{D}}^2$       & $-$     & $-$    &   
                        $(1600)^2$ & $  (871)^2$ &
                        $(1650)^2$ & $ (1000)^2$ &
                        $(1309)^2$     & $(1440)^2$ \\[1ex] 
$m_{L_3}^2$           & $$     & $  $    &
                        $(2470)^2$ & $ (2520)^2  $    &
                        $(1000)^2$ & $ (1100)^2$&
                        $ (351)^2$ & $(363)^2$ \\[1ex]
$m_{L_{1,2}}^2$       & $$     & $  $    &
                        $(2470)^2$ & $ (2520)^2$ &
                        $(1000)^2$ & $ (1100)^2$&
                        $ (351)^2$ & $(364)^2$ \\[1ex]
$m_{E_3}^2$           & $$     & $  $    &
                        $(2470)^2$ & $ (2387)^2 $    &
                        $(1000)^2$ & $  (977)^2$&  
                        $ (161)^2$ & $(160)^2$ \\[1ex]
$m_{E_{1,2}}^2$       & $$     & $  $    &   
                        $(2470)^2$ & $ (2387)^2$ &
                        $(1000)^2$ & $  (977)^2$&
                        $ (161)^2$ & $(164)^2$ \\[1ex]
$m_N^2$               & $-$     & $-$    &
                        $(2470)^2$ & $ (2500)^2$ &
                        $(1000)^2$ & $ (1040)^2$ &
                        $$     & $$ \\[1ex]
$m_{Q_{1,2}}^2$           & $$     & $  $    &
                        $(2470)^2$ & $ (2560)^2$ &
                        $(1000)^2$ & $ (1590)^2$&
                        $(1350)^2$ & $(1510)^2$ \\[1ex]
$m_{u_{1,2}}^2$           & $$     & $  $    &
                        $(2470)^2$ & $ (2630)^2$ &
                        $(1000)^2$ & $ (1580)^2$&
                        $(1310)^2$ & $(1470)^2$ \\[1ex]
$m_{d_3}^2$           & $$     & $  $    & 
                        $(2470)^2$ & $ (2520)^2  $    &
                        $(1000)^2$ & $ (1520)^2$ &
                        $(1310)^2$ & $(1440)^2$ \\[1ex]
$m_{d_{1,2}}^2$           & $$     & $  $    & 
                        $(2470)^2$ & $ (2520)^2$ &
                        $(1000)^2$ & $ (1520)^2$ &
                        $(1310)^2$ & $(1470)^2$ \\[1ex]
$m_{H_u}^2$           & $-$     & $-$    &   
                        $(2470)^2$ & $ (2410)^2$ &
                        $(1000)^2$ & $ (1040)^2$ &
                        $$     & $$ \\[1ex]
$m_{H_d}^2$           & $-$     & $-$    & 
                        $(2470)^2$ & $ (2522)^2$ &
                        $(1000)^2$ & $ (1100)^2$ &
                        $$     & $$ \\[1ex]
$m_{H_3}^2$           & $-$     & $-$    &
                        $(2470)^2$ & $ (2560)^2$ &
                        $(1000)^2$ & $ (11300)^2$ & 
                        $$     & $$ \\[1ex]
$m_{S_{1,2}}^2$       & $-$     & $-$    &  
                        $(2470)^2$ & $ (2500)^2$ &
                        $(1000)^2$ & $ (1040)^2$ & 
                        $$     & $$ \\[1ex]
$m_{D_{1,2}}^2$       & $-$     & $-$    &  
                        $(2470)^2$ & $ (2570)^2$ &
                        $(1000)^2$ & $ (1550)^2$ &
                        $$     & $$ \\[1ex]
$m_{\bar{D}_{1,2}}^2$ & $-$     & $-$    & 
                        $(2470)^2$ & $ (2520)^2$ &   
                        $(1000)^2$ & $ (1520)^2$ & 
                        $$     & $$ \\[1ex]
\end{tabular}
\end{center}
\caption{The soft SUSY breaking mass-squared parameters for the models
discussed. $m_{1,2}^2$ and  
$m^2_{Q_i},m^2_{u_i},m^2_{d_i},m^2_{L_i},m^2_{E_i}$ are the mass
parameters of the MSSM Higgs doublets and 
the MSSM matter fields, while $m_S^2$ is the mass parameter of the SM
singlet $S$. In the $E_6$ models there are also exotic quarks $D,\bar{D}$,
additional doublets $H_{u,d,3}$, and singlets $S_{1,2},N$, with soft
mass-squares given above (in self-evident notation). 
 \label{soft}} \end{table}

\begin{table}
\begin{center}
\begin{tabular}{|l|c|c|c|c|}
 & SF & $E_6$ (UA) & $E_6$ (NUA) & GMSB \\[1ex]
\hline
$M_{Z'}$                 & 1000 & 1720 & 1620 & 1120\\[0.5ex]
$\a_{Z-Z'}$              & $2\times10^{-3}$ & $2\times10^{-3}$ & 
                           $10^{-3}$ & $4\times10^{-3}$ \\[0.5ex]
\hline
$m_{h^0_1}$            &  115 &    62 &    79 &   125 \\[0.5ex]
$m_{h^0_2}$            &  657 &   772 &   348 &   998 \\[0.5ex]
$m_{h^0_3}$            & 1000 &  1720 &  1620 &  1090 \\[0.5ex]
$m_{A^0}$              &  662 &   773 &   342 &   993 \\[0.5ex]
$m_{H^{\pm}}$          &  657 &   774 &   349 &   994 \\[0.5ex]
\hline
$m_{\tilde{\c}^0_1}$     & 719     & $-$1690 & $-$1580 & $-$1120 \\[0.5ex]
$m_{\tilde{\c}^0_2}$     & 684     &  $-$870 &  $-$939 & $-$1060 \\[0.5ex]
$m_{\tilde{\c}^0_3}$     & 592     &    53 &    94 &   142 \\[0.5ex]
$m_{\tilde{\c}^0_4}$     & 578     &   103 &   183 &   266 \\[0.5ex]
$m_{\tilde{\c}^0_5}$     & 72     &   876 &   946 &  1060 \\[0.5ex]
$m_{\tilde{\c}^0_6}$     & 37     &  1745 &  1671 &  1120 \\[0.5ex]
$m_{\tilde{\c}^{\pm}_1}$ & 44     &   103 &   183 &   266 \\[0.5ex]
$m_{\tilde{\c}^{\pm}_2}$ & 620     &   876 &   945 &  1060 \\[0.5ex]
\hline
$m_{\tilde{t}_1}$        & 1068 &   805    &  1250    & 1390 \\[0.5ex]
$m_{\tilde{t}_2}$        & 1040 &   817    &  1260    & 1500 \\[0.5ex]
$m_{\tilde{c}_1}$        &      &  2680  &  1740 & 1550
\\[0.5ex]
$m_{\tilde{c}_2}$        &      &  2740  &  1740 & 1590 \\[0.5ex]
$m_{\tilde{u}_1}$        &      &  2680 &  1740 & 1550 \\[0.5ex]
$m_{\tilde{u}_2}$        &      &  2740  &  1740 & 1590 \\[0.5ex]
$m_{\tilde{b}_1}$        &      &   781    &  1240    & 1240
\\[0.5ex]
$m_{\tilde{b}_2}$        &      &  2460    &  1440  & 1500 \\[0.5ex]
$m_{\tilde{s}_1}$        &      &  2680 &  1750 & 1430 \\[0.5ex]
$m_{\tilde{s}_2}$        &      &  2460 &  1440 & 1590 \\[0.5ex]
$m_{\tilde{d}_1}$        &      &  2680 &  1750 & 1430 \\[0.5ex]
$m_{\tilde{d}_2}$        &      &  2460 &  1440 & 1590 \\[0.5ex]
$m_{\tilde{\tau}_1}$     &      &  2460  &   975  &   88 \\[0.5ex]
$m_{\tilde{\tau}_2}$     &      &  2510  &  1220  &  525 \\[0.5ex]
$m_{\tilde{\mu}_1}$      &      &  2470  &   975 &  111 \\[0.5ex]
$m_{\tilde{\mu}_2}$      &      &  2510  &  1220 &  522 \\[0.5ex]
$m_{\tilde{e}_1}$        &      &  2460  &   975 &  111 \\[0.5ex]
$m_{\tilde{e}_2}$        &      &  2510  &  1220 &  522 \\[0.5ex]
\end{tabular}
\end{center}
\caption{The mass and mixing angle of the additional neutral
gauge boson, tree level Higgs masses, neutralinos, charginos, and physical
sparticle masses (in GeV). 
Note that the spectrum of the first example (a toy model) is not 
realistic.
\label{physical}}
\end{table}

\end{document}